\documentclass[preprint,times,tighten]{aastex6}
\usepackage{amssymb,amsmath,graphicx,natbib,hyperref,titlesec,url}
\usepackage{color,enumitem}

\newenvironment{longdescription}
  {\begin{description}[style=unboxed]}
  {\end{description}}

\newcommand{\hi}          {\mbox{\rm H{\small I}}}

\newcommand{\msun}        {\mbox{\rm M$_\odot$}}

\titleformat{\section}[block]
{\normalfont\large\filcenter\sffamily}{}{.5em}{\bfseries}

\titleformat{\subsection}[block]
{\normalfont\sffamily}{}{.5em}{\itshape}

\shorttitle{Key Science Goals for the ngVLA}
\shortauthors{ngVLA Science Advisory Council}
\begin{document}

\begin{flushright}
{\bf ngVLA Memo \#19}\\
Document version November 27$^{\rm th}$, 2017
\end{flushright}

\title{Key Science Goals for the Next Generation Very Large Array (ngVLA):\\
Report from the ngVLA Science Advisory Council}

\author{Alberto Bolatto${}^1$, Shami Chatterjee${}^2$, Caitlin M.~Casey${}^3$,
	Laura Chomiuk${}^4$, Imke de~Pater${}^5$, Mark Dickinson${}^6$,
	James Di~Francesco${}^7$, Gregg Hallinan${}^8$, Andrea Isella${}^9$,
    Kotaro Kohno${}^{10}$, S.~R.~Kulkarni${}^{8}$, Cornelia Lang${}^{11}$,
    T.~Joseph~W.~Lazio${}^{12}$, Adam K.~Leroy${}^{13}$, Laurent Loinard${}^{14}$,
    Thomas J.~Maccarone${}^{15}$, Brenda C.~Matthews${}^{16}$, Rachel A.~Osten${}^{17}$,
    M.~J.~Reid${}^{18}$, Dominik Riechers${}^{2}$, Nami Sakai${}^{19}$,
    Fabian Walter${}^{20}$, \& David Wilner${}^{18}$\\
    \bigskip
    {\footnotesize\textnormal{%
    ${}^1${Univ.~Maryland}, ${}^2${Cornell Univ.}, ${}^3${Univ.~Texas at Austin}, 
    ${}^4${Michigan State Univ.}, ${}^5${Univ.~California, Berkeley}, ${}^6${NOAO}, 
    ${}^7${National Research Council of Canada}, ${}^8${California Institute of Technology}, ${}^9${Rice Univ.},
    ${}^{10}${Univ.~Tokyo}, ${}^{11}${Univ.~Iowa},
    ${}^{12}${Jet Propulsion Laboratory, California Institute of Technology}, ${}^{13}${Ohio State Univ.},
    ${}^{14}${Max Planck Institut f\"ur Radioastronomie}, ${}^{15}${Texas Tech Univ.},
    ${}^{16}${National Research Council of Canada}, ${}^{17}${Space Telescope Science Institute},
	${}^{18}${Harvard-Smithsonian Center for Astrophysics},
    ${}^{19}${Institue of Physical and Chemical Research (RIKEN)}, ${}^{20}${Max Planck Institute f\"ur Astronomie}
    }}}

\maketitle

\section*{Executive Summary}
%\begin{abstract}
This document describes some of the fundamental astrophysical problems that require observing capabilities at millimeter- and centimeter wavelengths well beyond those of existing, or already planned, telescopes. The results summarized in this report follow a solicitation from the National Radio Astronomy Observatory to develop key science cases for a future U.{}S.-led radio telescope, the “next generation Very Large Array” (ngVLA).  The ngVLA will have roughly 10 times the collecting area of the Jansky VLA, operate at frequencies from 1~GHz to 116~GHz with up to 20~GHz of bandwidth, possess a compact core for high surface-brightness sensitivity, and extended baselines of at least hundreds of kilometers and ultimately across the continent to provide high-resolution imaging. The ngVLA builds on the scientific and technical legacy of the Jansky VLA and ALMA, and will be designed to provide the next leap forward in our understanding of planets, galaxies, and black holes.

The ngVLA Science Advisory Council (ngVLA-SAC), a group of experts appointed by the NRAO, in collaboration with the broader international astronomical community, have developed over 70 compelling science cases requiring observations between 1~GHz and~116~GHz with sensitivity, angular resolution, and mapping capabilities far beyond those provided by the Jansky \hbox{VLA}, \hbox{ALMA}, and the SKA. These science cases span a broad range of topics in the fields of planetary science, galactic and extragalactic astronomy, as well as fundamental physics. The individual science cases were reviewed and discussed by the different science working groups within the ngVLA-SAC with the goal of distilling the top scientific goals for a future radio telescope. Below we outline the key science goals to come out of this process, in no particular order.  

\begin{longdescription}
\item[Unveiling the Formation of Solar System Analogues]%
The ngVLA will measure the planet initial mass function down to a mass of 5--10 Earth masses and unveil the 
formation of planetary systems similar to our own Solar System by probing the presence of planets on orbital radii as small as 0.5 AU at the distance of 140 pc. The ngVLA will also reveal circumplanetary disks and sub-structures in the distribution of mm-size dust particles created by close-in planets and will measure the orbital motion of these features on monthly timescales.
\item[Probing the Initial Conditions for Planetary Systems and Life with Astrochemistry]%
The ngVLA will be able to detect predicted, but as yet unobserved, complex prebiotic species that are the basis of our understanding of chemical evolution toward amino acids and other biogenic molecules. It will also allow us to detect and study chiral molecules, testing ideas on the origins of homochirality in biological systems. The detection of such complex organic molecules will provide the chemical initial conditions of forming solar systems and individual planets. 
\item[Charting the Assembly, Structure, and Evolution of Galaxies from the First Billion Years to the Present]%
The ngVLA will provide an order-of-magnitude improvement in depth and area for surveys of cold gas in galaxies back to early cosmic epochs, and it will enable routine sub-kiloparsec scale resolution imaging of the gas reservoirs. 
The ngVLA will afford a unique view into how galaxies accrete and expel gas and how this gas is transformed inside galaxies. It does so by imaging their extended atomic reservoirs and circum-galactic regions, and by surveying the physical and chemical properties of molecular gas over the entire local galaxy population. These studies will reveal the detailed physical conditions for galaxy assembly and evolution throughout the history of the universe.
\item[Using Pulsars in the Galactic Center as Fundamental Tests of Gravity]%
Pulsars in the Galactic Center represent clocks moving in the space-time potential of a super-massive black hole and would enable qualitatively new tests of theories of gravity. More generally, they offer the opportunity to constrain the history of star formation, stellar dynamics, stellar evolution, and the magneto-ionic medium in the Galactic Center.
The ngVLA combination of sensitivity and frequency range will enable it to probe much deeper into the likely Galactic Center pulsar population to address fundamental questions in relativity and stellar evolution.
\item[Understanding the Formation and Evolution of Stellar and Supermassive Black Holes in the Era of Multi-Messenger Astronomy]
The ngVLA will be the ultimate black hole hunting machine, surveying everything from the remnants of massive stars to the supermassive black holes that lurk in the centers of galaxies. High-resolution imaging abilities will allow us to separate low-luminosity black holes in our local Universe from background sources, thereby providing critical constraints on the formation and growth of black holes of all sizes and mergers of black hole-black hole binaries. The ngVLA will also identify the radio counterparts to transient sources discovered by gravitational wave, neutrino, and optical observatories. Its high-resolution, fast-mapping capabilities will make it the preferred instrument to pinpoint transients associated with violent phenomena such as supermassive black hole mergers and blast waves.

%\item[High Precision Astrometry to Determine the Size of the Galaxy and the Scale of the Universe]%
%With the significant increase in the collecting area of the ngVLA over the VLA, astrometric measurements using very large baseline techniques will reach a precision never before attained. Distances are fundamental to astronomy, and radio wavelength astrometry has consistently helped define the scale of our Universe --- from the nearest stars to distant galaxies. The capabilities of the \hbox{ngVLA}, at frequencies above 10~GHz and if equipped with continental baselines, would improve upon current astrometric capabilities (as well as extremely high resolution imaging of non-thermal sources) considerably.
\end{longdescription}
%\end{abstract}

\clearpage

\section{Unveiling the Formation of Solar System Analogues}

%\subsection{Driving Science Use Cases}
%Cradle of Life: PF3

%\subsection{Related Science Use Cases}
%Cradle of Life: PF1, PF5 

\subsection{Scientific Rationale}

Planets are thought to be assembled in disks around pre-main sequence stars but the physical processes responsible 
for their formation are poorly understood. Only recently, optical, infrared, and (sub-)millimeter telescopes have 
achieved the angular resolution required to spatially resolve the innermost regions of nearby protoplanetary disks. 
These efforts resulted in
the discovery of morphological features (rings, spirals, and crescents) in the distribution of circumstellar gas and
dust with characteristic sizes larger than 20 au \citep[e.g.][]{casassus13,vandermarel13,perez14,alma15,andrews16,isella16}. These structures are suggestive of gravitational perturbations of
yet unseen giant planets and provide a powerful tool to measure planet masses and orbital radii, study the circumplanetary environment, and investigate how forming planets interact with the circumstellar material \citep[e.g][]{jin16}. The angular resolution and sensitivity of current multi-wavelength disk imagery is limited to probing for the presence of planets more massive than Neptune at orbital radii larger than 20-30 au. A key science goal of the ngVLA will be to image the formation of super-Earths and giant planets across the entire disk, particularly within 10 au from the central star, and to probe for the presence of planets with masses as low as 5-10 Earth masses. The ngVLA will allow us to measure the initial mass and the birth radius functions of giant and massive rocky planets. Observations with the ngVLA will provide key information to understand the diverse demographics of exoplanetary systems and,  ultimately, unveil the formation of planetary systems similar to our own Solar system.

\begin{figure}[b]
\centering
\includegraphics[width=0.9\textwidth]{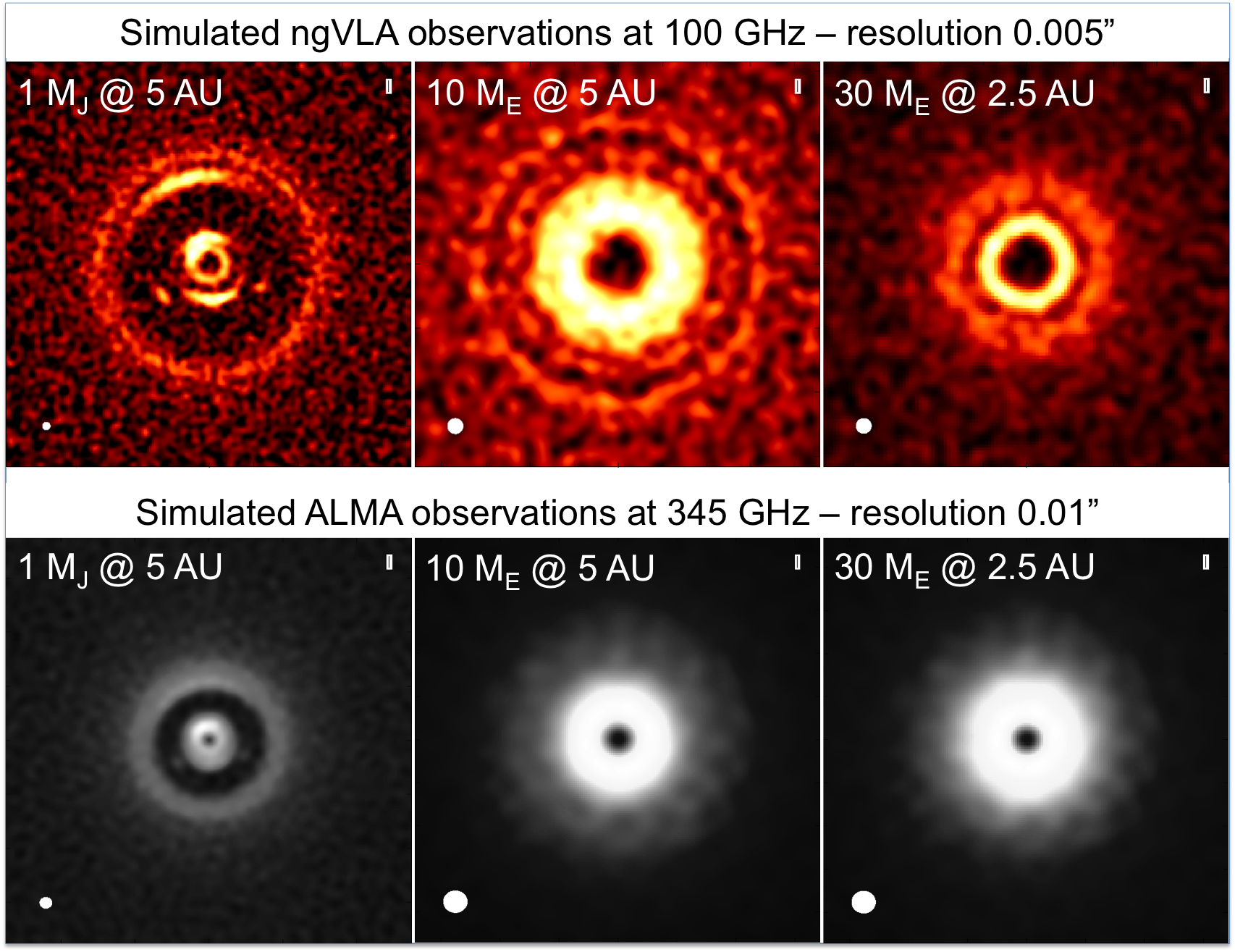}
\caption{Simulated observations (ngVLA on top row, ALMA on bottom row) of the continuum emission of a protoplanetary disk perturbed by a Jupiter mass planet orbiting at 5 au (left), a 10 Earth mass planet orbiting at 5 au (center), and a 30 Earth-mass planet orbiting at 2.5 au (right). The ngVLA observations at 100 GHz were simulated assuming an angular resolution of 5 mas and a rms noise level of 0.5 $\mu$Jy beam$^{-1}$. ALMA observations at 345 GHz where simulated assuming the most extended array configuration comprising baselines up to 16 km and a rms noise level of 8 $\mu$Jy beam$^{-1}$. From \citet{ricci18}.}
\end{figure}

Figure 1 illustrates the capabilities of the ngVLA for imaging planetary systems in the act of forming. The figure
compares simulated ngVLA observations at a frequency of 100 GHz to simulated ALMA observations at a frequency of
345 GHz, which provide the best compromise between angular resolution and sensitivity to the dust thermal emission.
Observations with the ngVLA will clearly reveal the presence of planets with masses as low as 10 Earth masses at orbital radii as
small as 2.5 au (central and right panels). These planets could not be detected by ALMA because of the high optical
depth of the dust emission at 345 GHz, and the lower angular resolution of its observations.  The ngVLA will be also superior to ALMA at imaging perturbations generated by giant planets orbiting close to the central 
star (left panels). In particular, the ngVLA will have the potential to detect circumplanetary disks and Trojan satellites around young Jupiter analogs. The exquisite angular resolution of the ngVLA will enable the measurement of the orbital motions of all these structures on monthly timescales, opening a completely new dimension to the study of  planet formation. 

At 30---50 GHz the ngVLA will probe the distribution of dust grains larger than those probed by ALMA, while avoiding the ionized gas emission that frequently dominates at lower frequencies \citep[e.g.,][]{ricci17}. These grains are expected to play a key role in the formation of planetesimals and planetary embryos. The combination of multi-wavelength ALMA and ngVLA observations will allow us to study the grain growth process, and probe the dynamical interaction between dust and gas particles before, during, and after planet formation.

\subsection{Telescope Requirements:}
Continuum observations between 20 GHz and 110 GHz with angular resolution better than 0.01\arcsec\ are required 
to study the formation of planets in the innermost 10 au of nearby ($<$140 pc) proto-planetary disks. 
Through extensive simulations of the disks perturbed by planets \citep[][Figure 1]{ricci18}, 
we calculate that a sensitivity of 0.5 $\mu$Jy beam$^{-1}$ in the continuum at 100 GHz is required to 
map structures in the dust distribution created by planets of mass down to 10 Earth-masses and orbital radius of 2.5 au. Observations would benefit from the largest possible aggregate bandwidth, and from polarization capabilities to constrain better the properties of the dust grains. A field of view larger than 2\arcsec\ would be required to map the entire disk in a single pointing. A maximum recoverable scale of at least $0\farcs5-1\arcsec$ would be required to minimize the effects of spatial filtering.

%\clearpage
\section{Probing the Initial Conditions for Planetary Systems and Life with Astrochemistry}

%\subsection{Driving Science Use Case}
%Cradle of Life: AC5

%\subsection{Related Science Use Cases}
%CoL: AC1, AC4, AC6, SS06; and AC2, AC3 (both with amended rms values)

\subsection{Scientific Rationale}
One of the most challenging aspects in understanding the origin and evolution of planets and planetary systems is tracing the influence of chemistry on the physical evolution of a system from a molecular cloud to a solar system.  The ngVLA will enable unprecedented observations of interstellar chemistry from the densest star-forming regions of the galaxy to protoplanetary disks. Existing facilities have already shown the stunning degree of molecular complexity present in these systems. The unique combination of sensitivity and spatial resolution offered by the ngVLA will permit the observation of both highly complex and very low-abundance chemical species that are exquisitely sensitive to the physical conditions and evolutionary history of their sources, which are out of reach of current observatories.  In turn, by understanding the chemical evolution of these complex molecules, unprecedentedly detailed astrophysical insight can be gleaned from these astrochemical observations.

For example, the thermal history of the densest regions of protoplanetary disks, which the ngVLA will be uniquely capable of imaging, can be studied by careful observation of the D/H ratio of organic molecules in that region. The chemical reactions which inject deuterium into these species, largely through reaction with H$_2$D$^+$, proceed more efficiently at lower temperatures.  This enhancement is known quantitatively and to high accuracy, and the resulting D/H ratio is well-preserved once temperatures are increased \citep[e.g.][]{cazaux11}.  Thus, observations of rare, low-abundance deuterated species provide quantitative insight into how cold the primordial mass reservoir was, and how long it stayed at these temperatures.  These species are often below the the detection threshold of modern facilities, especially in protoplanetary disks, but are well within reach of the ngVLA, particularly in the $70-110$~GHz range.

The ngVLA will also have sufficient sensitivity and angular resolution to resolve the ammonia snowline in protoplanetary disks.  Laboratory work suggests that ammonia may be the only viable molecular tracer of the water snowline, the most important chemical ``phase transition'' in a disk system from both the planetary-evolution and habitability perspectives \citep[e.g][]{salinas16}.  These same ammonia transitions, as well as complex molecules and rare isotopes, can also be observed in Solar System comets.  This in turn provides a more complete picture of (1) when these bodies where formed, (2) where in the natal disk system they were formed and have since migrated from, and (3) the level of inheritance of complex material from the pre-solar nebula. The level of heterogeneity in comet compositions is a test of the formation mechanism of planetesimals, and provides key constraints for temperature and chemical models of the pre-solar disk. 

Similarly, understanding the current day coupling of gas abundances, temperature, and dynamics in giant planet atmospheres requires observations of various molecular species, many observable only in the radio. The lower the frequencies observed, the deeper within planetary atmospheres can be probed. For example, the water lines from $18-45$~GHz probe the stratosphere and troposphere of giant planets. While JWST will execute some complementary science at 5 $\mu$m, the vastly superior resolution of the ngVLA means that radio observations will be much better for understanding cloud structures within these objects. Understanding the giant planets of our own solar system is essential to interpreting observations of exoplanets around other stars.

Finally, the ngVLA will enable an unprecedented view into complex organic (prebiotic) chemical evolution in the ISM.  Observation of a substantial number of predicted, but as yet undetected, complex prebiotic species are needed to truly understand chemical evolution toward amino acids and other biogenic molecules.  We are rapidly approaching the point of diminishing returns at which deep observations with ALMA and the GBT will no longer reveal new spectral lines, due to a combination of sensitivity limits and line-confusion at higher frequencies. Both problems can be solved by sensitive observations in the cm-wave regime. State-of-the-art models predict these molecules will display emission lines with intensities that are easily detectable with the ngVLA, but well below the current detectability thresholds of existing telescopes including ALMA, GBT, and IRAM. Figure 2 shows simulations of a representative set of the types of molecules whose discovery will be enabled by the ngVLA: N, O, and S-bearing small aromatic molecules, direct amino acid precursors, biogenic species such as sugars, chiral molecules, and, possibly amino acids themselves.

A highlight of the unique prebiotic science that will be made possible by the ngVLA is the study of chirality and its drivers, particularly the origin of homochirality in biological systems. Chiral molecules, that is, molecules whose mirror image is not identical to the original, are central to biological function.  Indeed, the mystery of homochirality, nature's use of only one of the mirror images in most biological processes, plays a central role in our quest to understand the origins of life, as well as being considered a nearly unambiguous biomarker.  There is no energetic basis for the dominance in life of one handedness of a chiral molecule over another, but rather, a slight excess was likely inherited at some point in the evolutionary process, and amplified by life. Given that material in planetary systems has been shown to be inherited from their parent molecular clouds, an excess of a particular handedness in that cloud may be the spark which drives homochirality in a certain direction. One possible route to generate a chiral excess is through UV-driven photodissociation of chiral molecules by an excess of left or right circularly polarized light. The ability not only to detect, but to image the abundance of chiral species at spatial scales commensurate with observations of circularly polarized light toward star-forming regions would be a giant leap forward. Using known, polarization-dependent photodissociation cross sections from laboratory studies, these observations would enable quantitative estimates of potential UV-driven excess. While such studies are well beyond the capability of existing observatories, they would be achievable with the ngVLA.  Chiral molecules, like other complex species detected earlier, are necessarily large, with propylene oxide, the only detected chiral species to date \citep{mcguire16}, being perhaps the only example simple enough for detection with existing facilities. The ngVLA will provide the sensitivity and angular resolution required to detect additional, biologically-relevant chiral species, such as glyceraldehyde.

\begin{figure}[h!]
\centering
\includegraphics[width=1\textwidth]{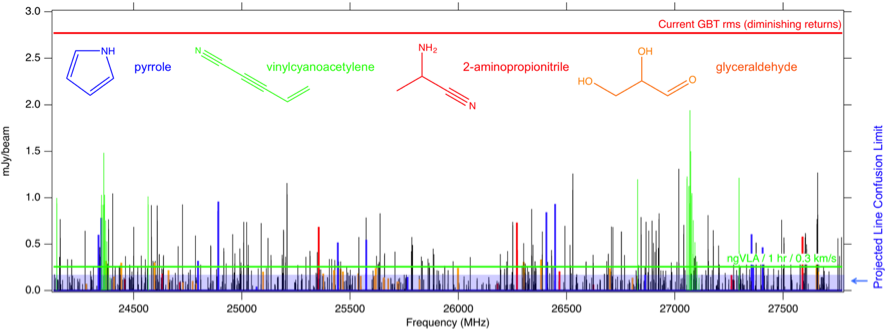}
\caption{A conservative simulation of a representative set of 30 currently undetected complex interstellar molecules (in black) which are likely to be detectable by the ngVLA above the confusion limit of an ngVLA survey. These lines are out of reach of current facilities. A few key molecules are highlighted in color. (Credit: B. McGuire)}
\end{figure}

\subsection{Telescope Requirements}
To meet the spectroscopic requirements to detect prebiotic molecules and complex organic molecules, the design of the ngVLA will need to reach rms levels of 30 uJy/beam\,km/s for frequencies between 16 GHz and 50 GHz with spectral resolution of 0.1 km/s. Angular resolution on the order of 50 mas is needed at 50 GHz and the largest angular scales expected range from $2\arcsec$ to $10\arcsec$.  The unique combination of sensitivity and resolution of the ngVLA will enable better depth than any pre-existing surveys for prebiotic molecules in the Galaxy. The deepest current survey for prebiotic molecules is being done with the GBT, but its spatial resolution limitations preclude any benefits from pushing any deeper in sensitivity. 

%\clearpage

\section{Charting the Assembly, Structure, and Evolution of Galaxies from the First Billion Years to the Present}  

%\subsection{Driving Science Use Cases}
%HiZ1, HiZ5, NGA2, NGA8

%\subsection{Related Science Use Cases}
%HiZ2, HiZ3, HiZ6, HiZ7, HiZ8, HiZ10, NGA5, NGA6, NGA7, NGA9, and NGA10

\begin{figure}[h!]
\centering
\includegraphics[width=0.8\textwidth]{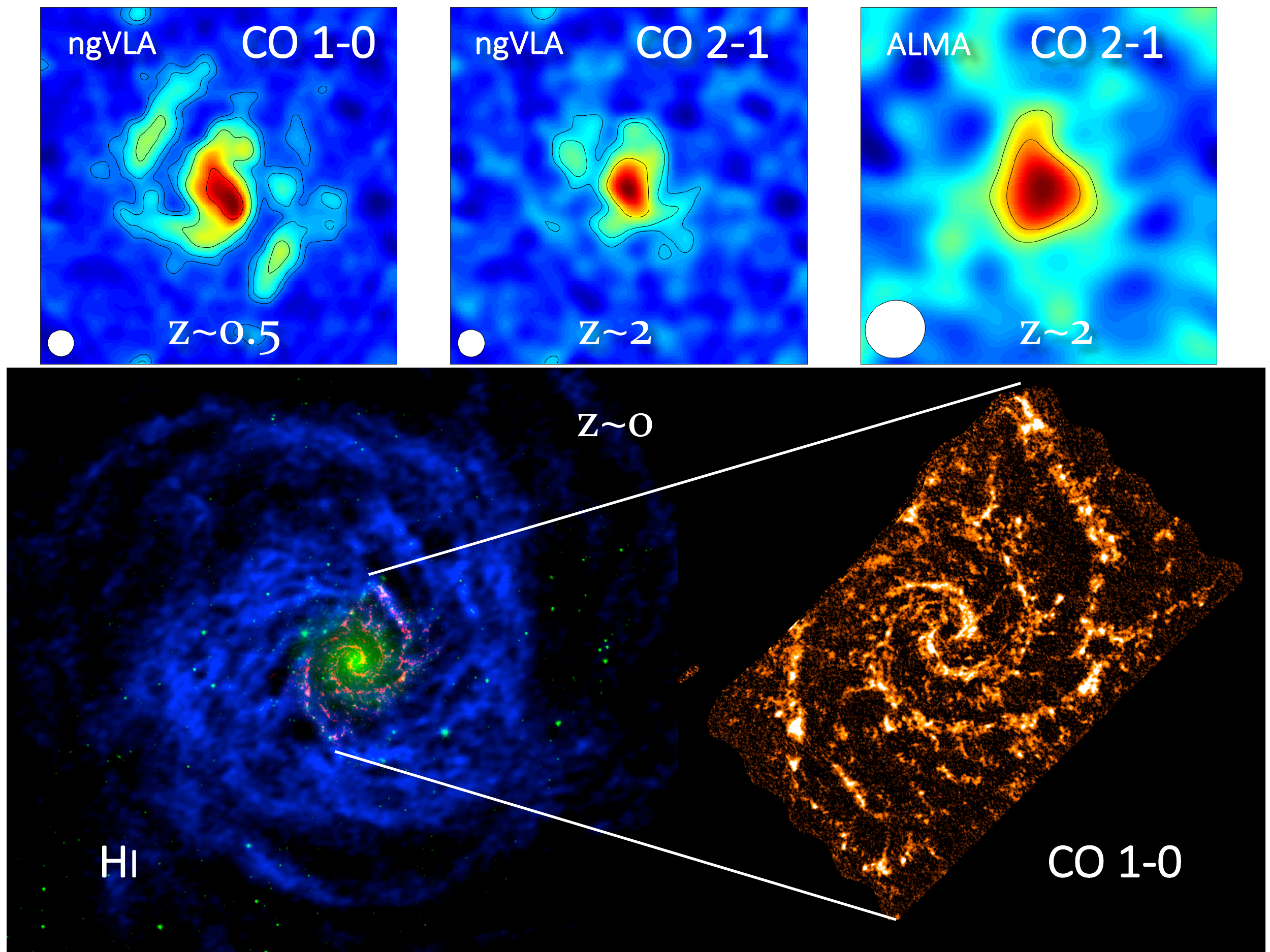}
\caption{The ngVLA will allow us to follow the evolution of galaxies from high redshift to the present day. The top panels illustrate the results of simulations based on M51 (the Whirlpool galaxy) with molecular mass scaled 
by factors of 1.4 ($z=0.5$) and 3.5 ($z=2$) to match it to the lowest molecular mass galaxies currently observable at ALMA and NOEMA \citep[from][]{memo13}. The corresponding SFR for the $z=2$ model would be 25\,M$_\odot$\,yr$^{-1}$. The synthesized beam shown in the bottom left corner is (left to right) $\theta=0.19\arcsec$, 0.20\arcsec, and 0.43\arcsec\ corresponding to spatial scales $L=1.2$, 1.7, and 3.7~kpc respectively. The corresponding maximum surface brightness is $T_B=6.7$, 2.1, and 1.0~K, and the black contours enclose regions with SNR$\geq$3, 5, 10, and 20.
The tapering of the beam is designed to provide the best compromise between angular resolution and S/N, and the integration times are 30 hours in all cases. The spatial and kinematic information recovered by the ngVLA allows the measurement of a precise rotation curve, which would only be possible to obtain from ALMA with an extremely large time investment. For full details see ngVLA memo \#13 \citep{memo13}.
The bottom panels show and example from the nearby universe: the grand-design spiral galaxy NGC~628 (M~74) located 7.3~Mpc away. This composite illustrates the molecular disk imaged in CO by ALMA (red; PHANGS survey, Schinnerer et al. in prep.), the stellar disk imaged at 4.5 $\mu$m by {\em Spitzer} \citep[green;][]{Kennicutt2003}, and the atomic disk  imaged in \hi\ by the VLA \citep[in blue;][]{Walter2008}, showing the atomic and molecular gas phases to which the ngVLA will be sensitive. The scale of this image is 15\arcmin\ in the vertical (North-South) direction, about 32~kpc at the distance of the object. The right panel shows a close-up of the area mapped in CO $J=2-1$ at 1\arcsec\ resolution: the ngVLA would be capable of quickly producing a significantly deeper map at similar resolution that would include not only CO $J=1-0$ but also dense gas tracers, molecular isotopologues, and many other molecules throughout the $\lambda=3-4$~mm band. The ngVLA will image the physical and chemical state and structure of the cold ISM at high resolution on statistically significant galaxy samples, providing unique insights into these processes.}
\end{figure}

\subsection{Scientific rationale} 
The processes that lead to the formation and evolution of galaxies throughout cosmic history involve the complex interplay between hierarchical merging of dark matter halos, accretion of primordial and recycled gas, transport of gas within galaxy disks, accretion onto central super-massive black-holes, and the formation of molecular clouds which subsequently collapse and fragment. The resulting star formation and black-hole accretion provide large sources of energy and momentum that not only light up galaxies but also bring about large changes in their gas reservoirs that we call feedback. How is gas accreted onto galaxies? What regulates the growth of galaxies through cosmic history? How is gas transported within galaxies and expelled by fountains and winds? How is gas inside galaxies influenced by local processes of star formation and black-hole accretion? How do the energetics, turbulent structure, self-gravity, density, and chemical state of  the gas change as the gas cycles between different phases, and how do these processes depend on galaxy properties or location in a galaxy? Observations of these processes not only constrain the dominant feedback mechanisms and timescales, but also establish useful chemical clocks and produce the observations necessary for interpreting spectroscopy across the universe out the highest redshifts. 

Our understanding of the processes that shape galaxies at cosmological distances is limited by our knowledge of the amount and physical properties of the cold gas available to fuel star formation. The ngVLA will be able to undertake large surveys to address systematically the molecular gas content and properties in high-$z$ galaxies using the red-shifted CO $J=1\rightarrow0$ and $J=2\rightarrow1$ transitions. The access to the lower level CO transitions removes the uncertainties related to the unknown excitation state of the gas, which introduce rapidly increasing biases for higher transitions \citep[e.g.,][]{Riechers11, Ivison11}. To date, detailed studies of the molecular gas that fuels star formation are limited to the most luminous main-sequence galaxies and starbursts during the peak epoch of cosmic star formation at $z\sim1$ to 3 \citep[e.g.,][]{Tacconi13, CW13}, when about half the stellar mass of the universe formed, while the ngVLA will be able to image gas-rich galaxies down to $\sim10^9$\,\msun\ in 24 hours. Statistics of molecular detections over a broader range of galaxy masses, together with vastly increased sensitivity to continuum emission arising from synchrotron, free-free, and cold dust at high-$z$, will provide unique insights into the evolution of galaxies through cosmic time. Moreover, with the power to carry out in-depth, high-resolution imaging studies of distant galaxies, the ngVLA will make it possible to image routinely and systematically the sub-kiloparsec scale distribution and kinematic structure of molecular gas in both normal main-sequence galaxies and large starbursts, an extremely challenging task for the present generation of telescopes \citep[e.g.,][]{Hodge12}.

The ngVLA has the capability to carry out unbiased, large cosmic volume surveys at virtually any redshift down to an order of magnitude lower gas masses than currently possible, thus exposing the evolution of gaseous reservoirs from the earliest epochs to the peak of the cosmic history of star formation. With the collecting area to detect and image small amounts of low-excitation molecular material, the ngVLA will open up the study of the formation, growth, and evolution of disks through the influx and accretion of material in the form of minor mergers (Figure 3). The ngVLA will provide a key component of the combined high-resolution multi-wavelength studies that will be undertaken with 30\,m--class optical telescopes, JWST, ALMA, and upcoming large NASA missions to image the process of galaxy assembly throughout the formative times of cosmic history.

The nearby universe, on the other hand, provides the ideal laboratory to study the mechanisms that drive some aspects of galaxy evolution. Galaxies continue to accrete material from their surroundings throughout their existence. This material gathers in their outer disks as \hi\ gas, constituting the largest gas reservoir in galaxies. The ngVLA will provide the combination of surface brightness sensitivity and resolution necessary to understand the physical makeup of these reservoirs (Figure 3). Most of the star formation activity, however, occurs in inner disks where the gas is denser and mostly in molecular form \citep{Leroy08}. Transport of gas from the outer reservoirs to the inner regions is a key and poorly understood feature of galaxy evolution. As the gas moves inward it becomes denser and gathers into molecular clouds, which disperse and reform during inter-arm passages. In the path of gas to stars, the formation of molecular clouds is likely one of the``bottlenecks'' that leads to regulation of star formation in galaxies \citep{Dobbs14}. Star formation occurs within portions of these clouds with an efficiency that is much lower than the na\"{\i}ve expectation of collapse in a free-fall time scale, due to feedback effects that are probably mediated by turbulence, likely injected on large scales \citep{McKee07}. Episodes of concentrated massive star formation drive fountains, where enriched gas is cycled through the galaxy halo, and even larger episodes such as starbursts or accretion onto central super-massive black holes can drive substantial galaxy winds. These phenomena constrain and regulate the stellar mass growth in galaxies, as well as injecting metals, dust, and energy into the circum-galactic and intergalactic medium \citep{Veilleux05}.

The ngVLA has the capability to survey the structure of the cold, star-forming interstellar medium at the parsec-resolution of star-forming units out to the Virgo cluster ($\sim20-30$\,Mpc distances). It will image not only CO but also a host of other molecular tracers with transitions that are $1-1.5$ dex weaker, providing a range of cold interstellar medium diagnostics and mapping the motion, distribution, and physical and chemical state of the gas as it flows in from the outer disk, assembles into clouds, and experiences feedback due to star formation or accretion into central super-massive black holes. The power of the ngVLA makes possible large systematic studies, taking advantage of chemical footprints that are beyond the limits of current capabilities to yield key insights on the processes that shape galaxies. Simultaneously, imaging of free-free and synchrotron continuum emission provides the full context of star formation and accretion activity. The deep imaging of the atomic gas in the outskirts of galaxies, the outer regions that constitute the interface between the inner star-forming disk and the cosmic web, will distinguish between settled extended \hi\ disks, compact high velocity clouds, merger remnants, and tidal dwarfs and other tidal features. 

\subsection{Telescope requirements}
An order-of-magnitude improvement in effective collecting area over the VLA at 1\,cm, and over ALMA at 3\,mm is critical to probe down to faint galaxy populations in the early universe, and enable high-resolution imaging of wide-spread, low surface brightness emission in normal galaxies. A large array ``core'' is essential to retain the necessary point source sensitivity, and not over-resolve faint galaxies. A large instantaneous bandwidth is key to carry out efficiently blind surveys of large cosmic volumes in a single observation. At the same time, the large bandwidth will provide routine access to molecular species different than CO, such as HCN, HCO$^+$, or N$_2$H$^+$. These species can then be studied for large samples across a wide range in galaxy luminosities and masses through stacking of CO-detected galaxies. Access to transitions of formaldehyde (5 GHz and 14 GHz), ammonia ($23-27$ GHz), methanol (particularly the 36 GHz masers), deuterated molecules ($\sim70$ GHz), and a host of dense gas tracers ($\sim 90$ GHz) besides CO (115 GHz) and \hi\ (1.4 GHz) provides key probes in the nearby universe.  With full continuum capabilities the ngVLA will also be able to obtain simultaneously measurements of the star formation rate from free-free and radio recombination line emission. Accurate recovery of flux for extended objects will require placing some of the collecting area on very short baselines. 

Neither ALMA nor the SKA Phase~1 have the power to carry out these observations. The SKA Phase~1 will not have the frequency coverage necessary to do the vast majority of this science. Although ALMA can tackle some of these observations on a few objects, it simply lacks the collecting area to do large samples. ALMA Band~1, for example, is roughly equivalent to the Jansky \hbox{VLA} performance at Q-band, and ALMA Band~3 is almost an order of magnitude less sensitive than the \hbox{ngVLA} at 90~GHz. Only the ngVLA will be able to study these processes on significant galaxy samples, distant or nearby.

%\clearpage

\section{Using Pulsars in the Galactic Center as Fundamental Tests of Gravity}\label{sec:tdcp.gcpsr}

%\subsection{Driving Science Use Case}\label{sec:tdcp.gcpsr.case}
%TDCP1

\begin{figure}[tb]
\begin{minipage}{0.54\textwidth}
\includegraphics[width=\textwidth]{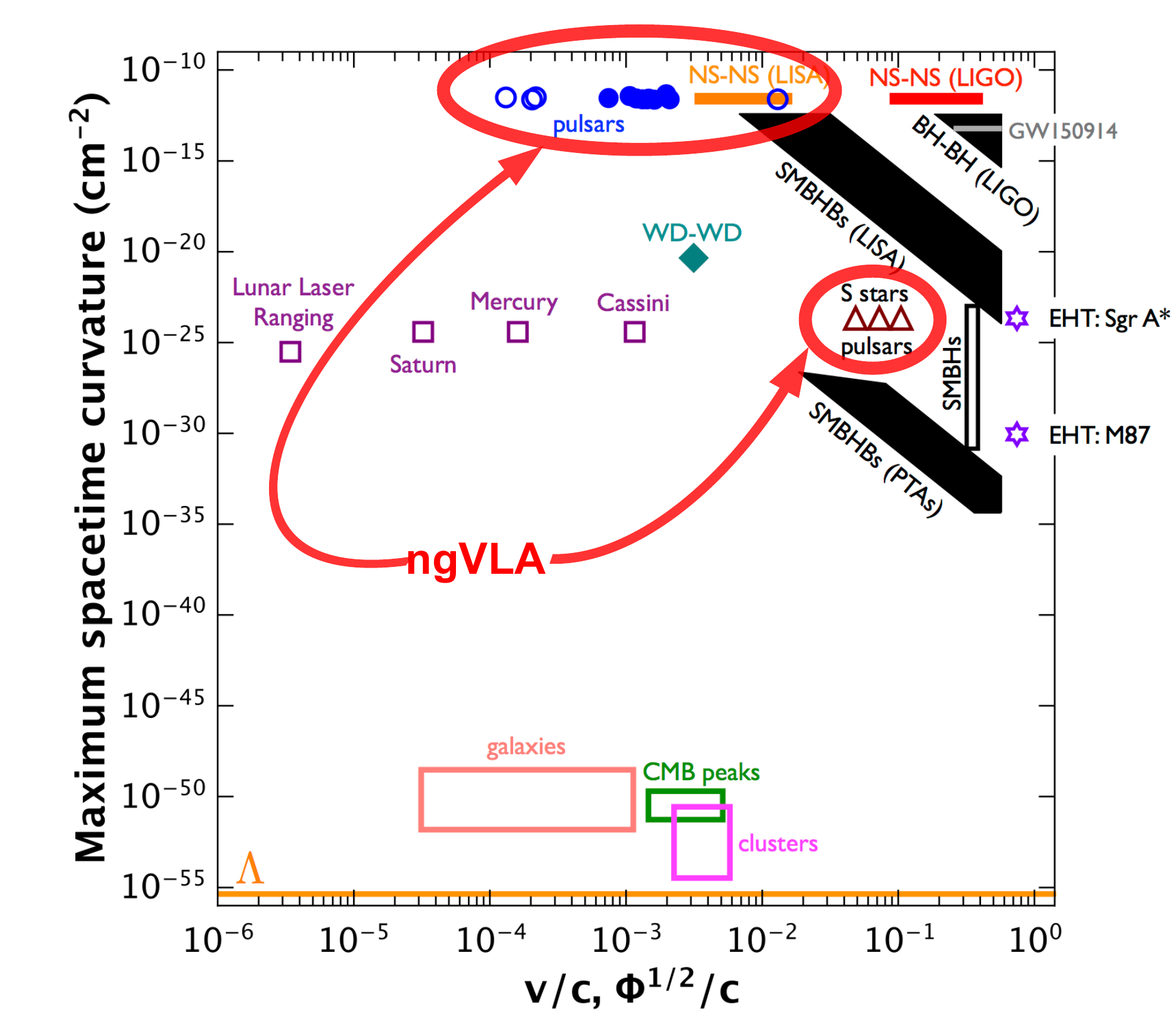}\hfil%
\end{minipage}\begin{minipage}{0.41\textwidth}
\includegraphics[width=\textwidth]{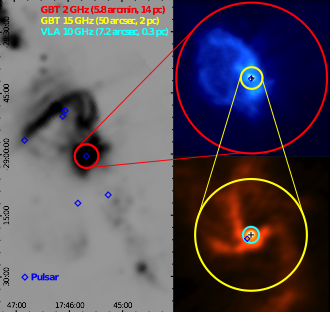}
\vspace{0.5cm}
\end{minipage}
\caption{(\textit{left}) Testing theories of gravity requires probing as close as possible to the strong field regime.  The abscissa shows the depth of the potential probed and the ordinate shows the spacetime curvature for the orbit of a test mass around a central mass---probing as far as possible into the upper right corner is most constraining on theories of gravity beyond General Relativity.  Pulsars near Sgr~A* probe regimes comparable to the infrared S~stars, but, as they represent clocks, pulsars probe different aspects of theories of gravity.  Pulsars in compact binaries, such as might form from three-body exchanges in the dense nuclear cluster, are in the upper center of the figure.  (Credit: M.~Kramer)
(\textit{right})  The distribution of known pulsars near the Galactic Center.  Despite being the region of highest density in the Galaxy and despite having been searched multiple times at a range of frequencies with sensitivities comparable to that of the \hbox{VLA}, only a small number of pulsars are known.  Even more puzzling, the closest pulsar to Sgr~A* is the magnetar PSR~J1745$-$2900, yet radio-emitting magnetars are an extremely rare sub-class amounting to less than $1\%$ of the field pulsar population. The combination of very high sensitivity and frequency coverage at $15-30$~GHz brought by the ngVLA will dramatically improve the ability to find pulsars in the Galactic Center. (Credit: R. Wharton)}
\label{fig:tdcp.gcpsr}
\vspace{-0.2cm}
\end{figure}

\subsection{Scientific Rationale}\label{sec:tdcp.gcpsr.science}
Pulsars in the Galactic Center represent clocks moving in the space-time potential of a super-massive black hole, which would enable qualitatively new tests of theories of gravity. More generally, pulsars offer the opportunity to constrain the history of star formation, stellar dynamics, stellar evolution, and the magneto-ionic medium in the Galactic Center. The high stellar densities in the Galactic Center region likely result in three-body interactions producing compact object binaries more extreme than those found in globular clusters, enabling hitherto impossible studies of General Relativity in a variety of sources.

There are a number of indications that the Galactic Center region contains a number of neutron stars. These range from the currently observed population of young, hot stars, to candidate pulsar wind nebulae and X-ray binaries, to estimates of the supernova rate derived from the diffuse X-ray emission. Millisecond pulsars in the inner Galaxy are the astrophysical alternative to dark matter annihilation to explain the observed gamma-ray {\em Fermi} excess. Relevant reviews on the expectations for millisecond pulsars in the Galactic Center include \citet{wharton12} and \citet{eatough15}. The resulting estimates for the number of active pulsars beamed toward the Earth are as high as 1,000.  Notably, given the possibility of exotic compact binaries and the strong space-time potential near Sgr~A*, it is possible for even canonical pulsars (with spin periods $\sim 1$~s) to provide useful measurements. 
%whereas, for many other kinds of pulsar studies, only millisecond pulsars are useful. 

Only a handful of pulsars in the central half-degree of the Galaxy are currently known.  Several factors make finding pulsars in the Galactic Center difficult. Pulsars are generally faint: pulsars at distances comparable to or greater than the distance to the Galactic Center represent only 20\% of the current census (480/2613), and nearly half of these have been discovered in the past 10~years even though pulsar searches have been conducted for half a century.  The steady increase in the discovery of more distant pulsars results from a combination of larger and more sensitive telescopes, larger bandwidth systems, and improved detection algorithms.  Moreover, not only are pulsars faint, but the intense Galactic emission toward the inner Galaxy increases the system temperature substantially at lower frequencies, leading to searches generally being less sensitive toward the inner Galaxy.  Finally, enhanced radio-wave scattering toward the inner Galaxy further decreases the effective sensitivity of searches, by increasing both the dispersion measure smearing and pulse broadening. Indeed, the first magnetar (highly magnetized pulsar) discovered near Sgr~A*, PSR~J1745$-$2900, shows substantial pulse broadening relative to most other pulsar lines of sight, although it is below original estimations. Observing at higher frequencies than are planned for the SKA can mitigate radio-wave scattering, but by itself the benefits are limited because of the generally steep radio spectra of pulsars. The ngVLA combination of a dramatic increase over present collecting area and the capability to observe at higher frequencies is necessary to find these objects (Figure 4).

Beyond the Galactic center, the ngVLA capabilities would enable another approach to probing gravity.  First, it is widely expected that the Galaxy may contain a small number of pulsar-black hole binaries \citep{lba05,okkb08}.  Likely even more powerful than relativistic neutron star-neutron star binaries for testing theories of gravity \citep{lewk14}, some pulsar-black hole binaries may form in the Galactic center region itself, due to three-body interactions resulting from the high stellar density in the region, in an analogous manner to how relativistic binaries are formed in globular clusters (which themselves may also host pulsar-black hole binaries).  Another formation channel, though, is through normal stellar evolution of a high-mass stellar binary.  Regardless of formation channel, if there are any in the Galaxy, they are likely to be rare and therefore distant (for example, in the current census of approximately 2,500 pulsars no pulsar-black hole binaries are known, though there are several neutron star-neutron star binaries).  Consequently, any pulsar-black hole binaries in the Galactic disk could experience significant pulse broadening, even if not as severe as that for lines of sight to the Galactic Center itself.  Moreover, a significant limitation to finding highly relativistic binaries could be the accelerations experienced by the pulsars.  The ngVLA imaging capabilities open up hybrid approaches to finding pulsars, conducting a search first for compact sources (potentially with steep spectra), followed by a targeted periodicity search on candidates. For example, \citet{bdfdijm17} used this hybrid imaging-periodicity technique in their recent successful detection of the recycled pulsar PSR~1751$-$2737.  The enhanced sensitivity of the ngVLA at radio frequencies of~3~GHz to~30~GHz will open a new door for the discovery and study of pulsars not only in orbit around Sgr~A*, but throughout the inner Galaxy.

\subsection{Telescope Requirements}\label{sec:tdcp.gcpsr.require}
The search for and study of Galactic Center pulsars is likely to drive the design of the ngVLA in three aspects. Frequency range: While there are uncertainties, and the distribution could be inhomogeneous, mitigating radio-wave scattering is likely to require a frequency range that includes the lower range anticipated for the ngVLA ($>3$\,GHz). Sensitivity:  Current searches with 100 m-class radio telescopes have found few pulsars, indicating that substantial additional sensitivity is necessary. An order-of-magnitude sensitivity would open the doors to new pulsar discoveries. Signal Processing: The scientific reward from pulsars results from precision timing, requiring a beam-forming capability.

%

%
%\clearpage
\section{Understanding the Formation and Evolution of Stellar and Supermassive Black Holes in the Era of Multi-Messenger Astronomy}\label{sec:tdcp.astro}

%\subsection{Driving Science Use Cases}
%TDCP2, TDCP5, TDCP7, TDCP8

%\subsection{Related Science Use Cases}
%NGA12

\subsection{Scientific Rationale}\label{sec:tdcp.astro.science}

\begin{figure}[t]
\centering
\begin{minipage}{0.6\textwidth}
\centering
 \includegraphics[width=0.8\textwidth]{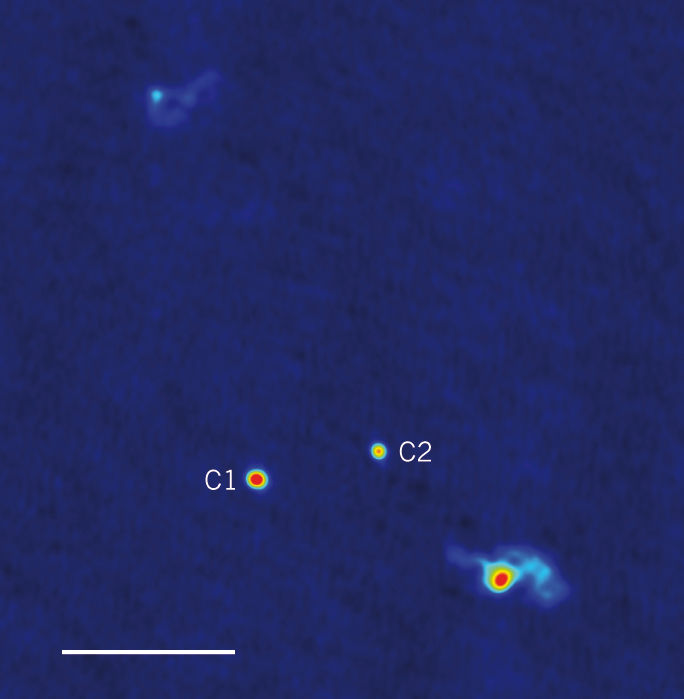}
 \end{minipage}\begin{minipage}{0.3\textwidth}
  \caption{The ngVLA will be an excellent tool for hunting black holes, including binary supermassive black holes. Here we show a binary system of SMBHs at $z = 0.06$. The black holes are separated by 7 pc (the white scale bar denotes 10 pc) with an orbital period of 30,000 yr, and jet emission is also observed extending from the black hole C2. The ngVLA, with its deep high resolution imaging capabilities, will enable discovery of many more such systems, with intimate synergies to LISA and Pulsar Timing Arrays. Image from \citet{Taylor2014}.}
\label{fig:taylor}
\end{minipage}
\end{figure}

While we now know that black holes exist on practically all mass scales, the astrophysics of how these objects form and grow remains a mystery. LIGO is now detecting black holes that are substantially more massive than previously known stellar mass black holes, and observing black hole--black hole mergers---although we do not know how black hole binaries form. While supermassive black holes (SMBHs) are thought to be widespread in galaxy centers, we do not understand how their growth was seeded or how (and how often) these extreme objects merge. The ngVLA, with its high sensitivity and high resolution, can answer all of these outstanding questions. 

%Moreover, as facilities such as aLIGO and IceCube are opening new astronomical windows through gravitational waves and neutrino observations, to progress further in our understanding of these phenomena we need to be able to localize and characterize the sources. Only the detection of the electromagnetic radiation associated with these energetic, and often cataclysmic events can provide precise localization, establish energetics and allow us to understand how such events interact with their surrounding environments. This is likely to remain challenging, even as other facilities come online over the next decade (Virgo, LISA, KAGRA, LIGO India, PINGU, IceCube Gen2, KM3NeT). Yet it is the combination of multi-messenger information that provides a complete picture of the life-cycle of massive stars, the micro-physics of their explosive deaths, and the formation and evolution of neutron stars, stellar black holes and super-massive black holes. What we know about the electromagnetic counterparts of gravitational waves and neutrino events and their frequency evolution indicates that the ngVLA will be a key component of the rapidly developing landscape of multi-messenger (electromagnetic-neutrino-gravitational wave) astronomy.

The ngVLA will enable a census of black holes on all scales, from stellar-mass to supermassive.  In the Milky Way Galaxy, the number of X-ray binaries (containing stellar mass black holes) is only weakly constrained to be somewhere in the range $10^2$--$10^8$ \citep{Tetarenko2016}, based on a small sample of just 20--30 known stellar-mass black holes \citep{McClintock2006}. Unaffected by dust obscuration and with the angular resolution to separate Galactic sources from background objects using proper motions, the ngVLA will be able to survey the Galaxy to detect jet-powered synchrotron emission from weakly accreting black holes and increase the black hole sample by at least an order of magnitude.  Simply measuring the size of the population would have profound implications for key parameters impacting binary black hole formation, such as common envelope evolution and the strength of dynamical ``kicks'' delivered to black holes at birth (caused by asymmetries in the parent core-collapse supernovae). These parameters are key inputs into one of the core problems that has already developed in gravitational wave astronomy --- whether double black holes form through normal binary stellar evolution, or whether they require globular cluster formation mechanisms. The ngVLA will also directly measure black hole natal kicks through determination of proper motion and parallax. Finally, the ngVLA will be a superb tool for multi-wavelength follow-up of these discoveries to measure the black hole mass distribution. The ngVLA is uniquely positioned for black hole survey science, as black holes in the Galaxy will be scatter-broadened at lower frequencies, precluding high-resolution imaging. In addition, we expect the radio spectra of low-luminosity accreting black holes to be flat ($S_{\nu} \sim \nu^0$). The ngVLA wide field of view will enable surveys that are simply not possible at higher frequencies.

While SMBHs dwell at the centers of many, if not all galaxies, we still do not understand how black holes manage to grow to masses, $10^6$--$10^{10}$~M$_{\odot}$. Lead contender models include the merger of less massive black hole ``seeds'' (i.e., remnants of Population III stars) and direct collapse of more massive black holes in early dark matter halos \citep{Volonteri2003}. These models can be best tested by measuring the occupation fraction of SMBHs in nearby, low-luminosity galaxies.
%where the gravitational radiation rocket effect  could eject black holes at the time of merger, perhaps leaving many of the lower mass galaxies without central supermassive black holes \citep{Redmount1989}. 
Deep, high-resolution imaging of nearby galaxies will provide proper motion distinction between nearby low-luminosity active galactic nuclei and background sources ($\sim$10 $\mu$as yr$^{-1}$). Thereby, the ngVLA will enable the best search possible for SMBHs, measure their occupation fraction in dwarf galaxies, and test models of SMBH formation. 

By enabling both pulsar timing arrays and high-resolution imaging, the ngVLA will survey the population of binary SMBHs, measure the rate of super-massive black hole mergers through their contributions to the stochastic gravitational wave background, and image the evolution of binary SMBHs in the lead up to merger. Recently, \citet{Bansal2017} made the first measurements of the orbit of a SMBH binary with the VLBA (Figure \ref{fig:taylor}). With greater sensitivity, there will be a far greater chance to identify and measure the proper motions of double AGN in much tighter orbits, where precision tests of General Relativity could be made. Meanwhile, the ngVLA will be a key facility for monitoring millisecond pulsars to detect and characterize the nanohertz gravitational wave background from SMBH mergers, and will be complementary to gravitational wave observatories in the kHz and mHz bands.

The ngVLA will be particularly important as a follow-up tool for the $10-100$s of coalescing binary SMBHs detected by LISA each year. A general prediction of General Relativistic MHD simulations of coalescing binary SMBHs 
%surrounded by a toroidal magnetic field 
is the occurrence of a 
%transient, 
collimated Poynting-flux outflow at the moment of coalescence, which may exceed the Eddington luminosity of the system. Such an outflow is likely to be observable as a few-day to few-week radio transient peaking at $\nu\gg10$~GHz and exceeding 100~$\mu$Jy at a redshift $z\sim6$. The ngVLA, with its high-frequency capabilities and outstanding survey speed will be ideally suited to discover and localize such transients within the 
%$\sim 10$ square degree 
error regions of LISA, and will be instrumental in unlocking the scientific potential of gravitational-wave events detected by LISA.

The ngVLA will come on-line at the culmination of a phase of rapid growth in gravitational wave and neutrino astronomy capabilities (e.g., Virgo, LISA, KAGRA, LIGO India, PINGU, IceCube Gen2, KM3NeT).
%, and coinciding with the mature phase of the Large Synoptic Survey Telescope (LSST), the deepest probe of the optical transient sky yet conceived. 
To make further progress in our understanding of these phenomena we need to be able to localize and characterize these multi-messenger (gravitational wave-neutrino-electromagnetic) sources. Only the detection of the electromagnetic radiation associated with these energetic, and often cataclysmic events can provide precise localization, establish energetics and allow us to understand how such events interact with their surrounding environments. It is the combination of multi-messenger information that will provide a complete picture of the life-cycle of massive stars, the micro-physics of their explosive deaths, and the formation and evolution of stellar and super-massive black holes. 
The ngVLA will add top-notch capabilities in terms of sensitivity, survey speed, localization capability and frequency agility. The latter is particularly important, as the synchrotron spectra of explosive transients reflect the energetics of the ejected mass and the density of the surrounding interstellar medium. With the ability to observe across a range of frequencies from 1 GHz to 116 GHz, the ngVLA will be able to target the full gamut of explosive events occurring in a range of diverse environments, ranging from neutron star mergers (peaking at $\sim$1.5 GHz) to mergers of SMBHs (peaking at $\nu\gg10$~GHz). What we know about the electromagnetic counterparts of gravitational wave and neutrino events and their frequency evolution indicates that the ngVLA will be a key component of the rapidly developing landscape of multi-messenger astronomy.

%Moreover, the ngVLA will be able to target the full gamut of explosive events occurring in a range of diverse environments, ranging from neutron star mergers to SMBH-mergers. The ability to pursue the necessary high frequency observations sets the ngVLA apart from the SKA: supernovae, GRBs, and TDE all peak a frequencies that are higher than those that will be observed by the SKA. 

\subsection{Telescope Requirements}\label{sec:tdcp.astro.require}

The high resolution capabilities of the ngVLA are crucial for surveying black holes. 
%as the Galactic census of stellar mass black holes must mitigate confusion in the Galactic plane and the nearby dwarf galaxy census of supermassive black holes must be able to localize and constrain the proper motion of low-luminosity systems.  
High resolution imaging will enable proper motion separation of local black holes (both Galactic and in nearby galaxies, out to $\sim$15 Mpc) from background sources. Long baselines will also enable the ngVLA to image the SMBH binaries that will be detected in gravitational waves by LISA and pulsar timing arrays. %Although many of these goals are possible with baselines up to $300-500$ km, 
%Proper-motion separation of local black holes from background sources 
These astrometric science goals benefit from the implementation of very long baselines ($\sim$1000\,km for mas--$\mu$as accuracy). While the key frequency range is $5-20$~GHz, the availability of higher frequencies will be critical in regions with high interstellar scatter broadening. Pulsar timing array observations require sub-array capabilities (ideally at least 5 sub-arrays) and the coverage of frequencies down to $1-2$~GHz.

Gravitational wave detectors are expected to provide localization within a $7-10$ square degree area, when LISA is operational. High survey speed, covering a localization region to the appropriate depth, is a key requirement. This involves mapping a 10 square degree region to a depth of $\sim 1$ $\mu$Jy for detection of NS-NS and NS-BH mergers, requiring $\lesssim 10$ hours of on-the-fly mapping per epoch of each event for the current ngVLA notional Band 1 parameters ($\sim 2.2$ GHz with 2.7 GHz bandwidth). Similarly, the detection of LISA-detected SMBH mergers involves mapping a 10 square degree region to a depth of $\sim 10$ $\mu$Jy, requiring $\lesssim 10$ hours of on-the-fly mapping per epoch of each event for the current ngVLA Band 4 notional parameters ($\sim 27$ GHz with 14 GHz bandwidth). The ability to receive and respond to external triggers is also an essential requirement to enable this science.

\paragraph*{Acknowledgements}
The National Radio Astronomy Observatory is a facility of the National Science Foundation operated under cooperative agreement by Associated Universities, Inc.
Part of this work was carried out at the Jet Propulsion Laboratory, California Institute of Technology, under contract with the National Aeronautics and Space Administration.

\end{document}